\documentclass[traditabstract,longauth]{aa}
\usepackage{graphicx}
\usepackage{txfonts}

\begin{document}
 \title{The \emph{Herschel}\thanks{{\it \emph{Herschel}} is an ESA space observatory with science instruments provided by European-led Principal Investigator consortia and with important participation from NASA.} view of the on-going star formation in the Vela-C molecular cloud}
\author{T. Giannini\inst{\ref{oar}}
		 \and
          D. Elia\inst{\ref{ifsi}}
          \and
          D. Lorenzetti\inst{\ref{oar}}
          \and 
          S. Molinari\inst{\ref{ifsi}}
		  \and
 		  F. Motte\inst{\ref{paris}}
		  \and
          E. Schisano\inst{\ref{ifsi}}
          \and
          S. Pezzuto\inst{\ref{ifsi}}
		  \and
          M. Pestalozzi\inst{\ref{ifsi}}  
		  \and
          A. M. Di Giorgio\inst{\ref{ifsi}}
		  \and
          P. Andr\'e\inst{\ref{paris}}
	      \and
          T. Hill\inst{\ref{paris}}
          \and 
          M. Benedettini\inst{\ref{ifsi}}
          \and
          S. Bontemps\inst{\ref{bord}}
		  \and
 		  J. Di Francesco\inst{\ref{can1},\ref{can2}}
  		  \and
 		  C. Fallscheer\inst{\ref{can1},\ref{can2}}
          \and
		  M. Hennemann\inst{\ref{paris}}
		  \and
          J. Kirk\inst{\ref{cardiff}}
          \and 
          V. Minier\inst{\ref{paris}}
          \and
          Q. Nguy$\tilde{\hat{\rm e}}$n Lu{\hskip-0.65mm\small'{}\hskip-0.5mm}o{\hskip-0.65mm\small'{}\hskip-0.5mm}ng\inst{\ref{sac}}
 		  \and
          D. Polychroni\inst{\ref{ifsi}}
          \and 
          K.L.J. Rygl\inst{\ref{ifsi}}
          \and 
          P. Saraceno\inst{\ref{ifsi}}
          \and 
          N. Schneider\inst{\ref{paris}}
          \and
          L. Spinoglio\inst{\ref{ifsi}}
          \and 
		  L. Testi\inst{\ref{eso}}
          \and
		  D. Ward-Thompson\inst{\ref{uk}}
		  \and
          G. J. White\inst{\ref{open},\ref{mcmaster}}
          }
\institute{INAF-Osservatorio Astronomico di Roma, via Frascati 33, I-00040 Monte Porzio Catone, Italy\label{oar}
   \and
INAF-Istituto di Astrofisica e Planetologia Spaziale, via Fosso del Cavaliere 100, 00133 Rome, Italy\label{ifsi}
  \and
Laboratoire AIM, CEA/IRFU CNRS/INSU Universit\'e Paris Diderot, CEA-Saclay, 91191 Gif-sur-Yvette Cedex, France\label{paris}
\and
Universit\'e de Bordeaux, OASU, Bordeaux, France\label{bord}
\and
University of Victoria, Department of Physics and Astronomy, PO Box 3055, STN CSC, Victoria, BC, V8W 3P6, Canada\label{can1}
\and
National Research Council of Canada, Herzberg Institute of Astrophysics, 5071 West Saanich Rd., Victoria, BC, V9E 2E7, Canada\label{can2}
\and
School of Physics and Astronomy, Cardiff University, Queens Buildings, The Parade, Cardiff, CF243AA, UK\label{cardiff}
\and
Laboratoire AIM Paris-Saclay, CEA/IRFU - CNRS/INSU - Universit\'e Paris Diderot, Service d'Astrophysique, B\^at. 709, CEA-Saclay, F-91191 Gif-sur-Yvette Cedex, France\label{sac}
\and
ESO, Karl Schwarzschild str. 2, 85748 Garching bei Munchen, Germany\label{eso}
\and
School of Physics and Astronomy, Cardiff University, Queens Buildings, The Parade, Cardiff, CF243AA, UK\label{uk}
\and
Department of Physics and Astronomy, The Open University, Milton Keynes, UK\label{open}
\and
Department of Physics and Astronomy, McMaster University, Hamilton, ON, L8S 4M1, Canada\label{mcmaster}
 }   
\date{}

\abstract{
{{\it Aims.  } As part of the \emph{Herschel} guaranteed time key program 'HOBYS', we present the \emph{PACS} and \emph{SPIRE} photometric survey of the star forming region Vela-C, one of the nearest sites of low-to-high-mass star
formation in the Galactic plane. Our main objectives are to take a census of the cold sources and to derive their mass distribution down to a few solar masses.}\\
{{\it Methods. } Vela-C has been observed with \emph{PACS} and \emph{SPIRE} in parallel mode at five wavelengths between 70\,$\mu$m and 500\,$\mu$m over an area of about 3 square degrees. A photometric catalogue has been extracted from the detections in each of the five bands, using a  threshold of 5 $\sigma$ over the local background. Out of this catalogue we have selected a robust sub-sample of 268 sources, 
of which $\sim$\,75\% are cloud clumps (diameter between 0.05\,pc and 0.13\,pc) and 25\% are cores (diameter between 0.025\,pc and 0.05\,pc). Their Spectral Energy Distributions (SEDs) have been fitted with a modified black body function. We classify 48 sources as protostellar, based on their detection at 70\,$\mu$m or at shorther wavelengths, and 218 as starless, because of non-detections at 70\,$\mu$m. For two further sources, we do not provide a secure classification, but suggest they are Class\,0 protostars.} \\
{{\it Results. } From SED fitting we have derived key physical parameters (i.e. mass, temperature, bolometric luminosity). Protostellar sources are in general warmer ($\langle T \rangle$=12.8\,K) and more compact ($\langle$diameter$\rangle$=0.040\,pc) than starless sources ($\langle T \rangle$=10.3\,K, $\langle$diameter$\rangle $=0.067\,pc). Both these evidences can be ascribed to the presence of an internal source(s) of moderate heating, which also causes a temperature gradient and hence a more peaked intensity distribution. Moreover, the reduced dimensions 
of protostellar sources may indicate that they will not fragment further. A virial analysis of the starless sources gives an upper limit of 90\% for the sources gravitationally bound and therefore prestellar in nature. A luminosity vs. mass diagram of the two populations shows that protostellar sources are in the early accretion phase, while prestellar sources populate a region of the diagram where mass accretion has not started yet. 
We fit a power law $N$(log$M$) $\propto$ $M^{-1.1\pm0.2}$ to the linear portion of the mass distribution of prestellar sources. This is in between that typical of CO clumps and those of cores in nearby star-forming regions. We interpret this as a result of the inhomogeneity of our sample, which is composed of comparable fractions of clumps and cores.}

}
\authorrunning{T.Giannini et al.}
\titlerunning{The \emph{Herschel} view of the on-going star formation in Vela-C}
\maketitle
%

\section{Introduction}\label{sec:sec1}
Most of the stellar content in our Galaxy forms in cold ($T$\,$\sim$\,10 - 30\,K) and dense ($n$\,$>$\,10$^3$\,cm$^{-3}$) cores within the Giant Molecular Clouds (GMC) of the Galactic Disk (e.g., Blitz 1991). The main properties of the stellar population, such as the efficiency of the mass conversion into stars and the shape of the initial mass function (IMF) may very well be closely related to the physical properties and mass distribution of the progenitor structures in the parental cloud (e.g. Mc\,Kee \& Ostriker 2007).

Being one of the nearest GMCs in the Galactic disk, the Vela Molecular Ridge (VMR, Murphy \& May 1991, $l \approx$ 260$^{\circ}$-275$^{\circ}$, $b \approx \pm$3$^{\circ}$), represents an ideal observational target. It is composed 
of four  molecular clouds at a distance between 700 pc (clouds A,\,C,\,D) and 2000 pc (cloud B, Liseau et al. 1992), and harbours protostars 
with masses up to 10 M$_{\odot}$, both isolated and clustered (Massi et al. 2000, 2003). 
Vela-C is the most massive component and hosts the youngest stellar population (Yamaguchi et al. 1999); this latter has been investigated by combining near-infrared data ($J$,\,$H$,\,$K$-band images) with far-infrared surveys ($MSX$,\,$IRAS$) and around thirty isolated protostars and seven embedded young clusters associated with C$^{18}$O clumps have been found (Liseau et al. 1992; Lorenzetti et al. 1993; Massi et al. 2003, Baba et al. 2006). A bright HII region, RCW\,36, has been associated by Massi et al. (2003) with an early-type star (spectral type O5-B0). A second HII region, RCW\,34, was originally associated to Vela-C by several authors (Herbst 1975, Murphy \& May 1991), but recent observations favour a much longer distance ($d$=2.5 kpc, Bik et al. 2010).
Vela-C has been imaged at 250\,$\mu$m, 350\,$\mu$m and 500\,$\mu$m with the Balloon-borne Large Aperture Submillimeter Telescope (\emph{BLAST}, Pascale et al. 2008) that provided the first census of the compact dust emission in a range of evolutionary stages and lifetimes (Netterfield et al. 2009). 

As a target of the \emph{Herschel} guaranteed time key program  HOBYS ('\emph{Herschel} imaging survey of OB Young Stellar objects', Motte et al. 2010), Vela-C has been observed with the \emph{PACS} (Poglitsch et al. 2010) and \emph{SPIRE} (Griffin et al. 2010) cameras between 70\,$\mu$m and 500\,$\mu$m. The extended emission in the form of filaments and ridges has been presented by Hill et al. (2011) who identify five different sub-regions with different column densities. Here we will focus on the determination of the physical parameters of the compact sources, their evolutionary stage and their mass distribution. The latter has been investigated in several star forming regions through different tracers (e.g. Testi \& Sargent 1998, Motte et al. 1998, 2001, Kramer et al. 1998, Reid \& Wilson 2006, Enoch et al. 2008, Rathborne et al. 2009, K\"{o}nyves et al. 2010, Ikeda \& Kitamura 2011). 
In this paper, we intend to enlarge the statistics providing the mass distribution over a wide range, going from subsolar values up to tens 
of solar masses. 

This paper is organized as follows. In Section\,\ref{sec:sec2} we describe the observations, the data reduction procedures and the photometric results. In Section\,\ref{sec:sec3} we describe the Spectral Energy Distribution (SED) analysis along with the fitted physical parameters. In Section\,\ref{sec:sec4} we discuss the evolutionary stages of the sources along with their mass distribution. Our results are then summarized in 
Section\,\ref{sec:sec5}.


\section{Observations, data reduction and results}\label{sec:sec2}

\begin{figure*}
\centering
\includegraphics[width=14cm]{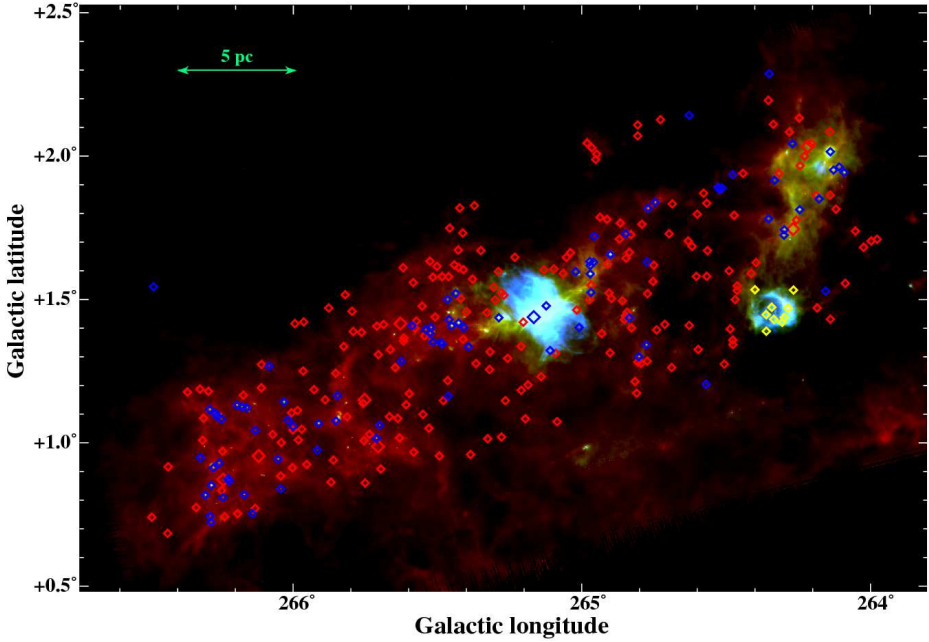}
\caption{Composite 3-color image of Vela-C : \emph{PACS} 70\,$\mu$m (blue), \emph{PACS} 160\,$\mu$m (green), and \emph{SPIRE} 500\,$\mu$m (red). The blueish regions near the center and towards the bottom right of the map are the HII regions RCW\,36 and RCW\,34, respectively. Red and blue diamonds represent the locations of prestellar and protostellar sources, respectively, belonging to the sample of 268 objects used for the SED analyisis. Bigger diamonds indicate sources with M\,$\ge$\,20\,M$_{\odot}$, while yellow diamonds indicate a few sources inside RCW\,34 (see text). The top left segment corresponds to a length of 5 pc in the sky for a distance to Vela-C of 700 pc.}
\label{tricro}
\end{figure*} 

Vela-C was observed on May 18, 2010 with the \emph{Herschel} parallel mode, i.e., using simultaneously \emph{PACS} at 70/160 $\mu$m and \emph{SPIRE} at 250/350/500 $\mu $m. A common area of $\sim$ 3 square degrees was covered by both the instruments around the position $\alpha_{(J2000.0)}$=08$^{\mathrm{h}}$59$^{\mathrm{m}}$55$^{\mathrm{s}}$, $\delta_{(J2000.0)}$=-43$^{\circ}$53$^{\prime}$00$^{\prime\prime}$. The field was observed in two orthogonal directions at the scan speed of $20\hbox{$^{\prime\prime}$}$/s. 
The data reduction strategy is described in detail in Traficante et al. (2011): here we summarize only the fundamental steps. From archival data to the Level 1 stage, we used scripts prepared in the \emph{Herschel} Interactive Processing Environment (HIPE, Ott 2010), partially customized compared to the standard pipeline. The obtained time ordered data (TODs) of each bolometric detector were then processed further by means of dedicated IDL routines and finally maps were created using the FORTRAN code ROMAGAL. Images of Vela-C at all wavelengths have been presented by Hill et al. (2011), here for the readers' convenience, we show in Figure \ref{tricro} the composite 3-color image at 70/160/500 $\mu$m.                                 

\subsection{Source Detection and Photometry}\label{sec:sec2.1}
The detection and photometry of compact sources have been carried out using the Curvature Threshold Extractor package (CuTeX, Molinari et al. 2011) applying the same strategy as Molinari et al. (2010). Briefly, in each band, sources were identified as peaks in the second-derivative images of the original HOBYS maps, then an elliptical Gaussian fit was performed to provide: 1) the total flux, integrated down to the zero intensity level, 2) the observed FWHM (defined as the geometrical mean of the major and minor ellipse axes), and 3) the peak intensity. The latter, when divided by the local rms noise, allows us to obtain an {\it a posteriori} estimate of the $S/N$ ratio. We filtered out all sources with $S/N<5$.
Following Elia et al. (2010), entries at different wavelengths in the \emph{PACS}/\emph{SPIRE} catalogue have been attributed to the same source based on simple positional criteria. In practice, we associate two sources detected in two different bands if their mutual angular distance does not exceed the radius of the \emph{Herschel} half-power beam-width (HPBW\footnote{The values of the \emph{Herschel} HPBW are 5.0$^{\prime\prime}$ at 70\,$\mu$m, 11.4$^{\prime\prime}$ at 160\,$\mu$m, 17.8$^{\prime\prime}$
at 250\,$\mu$m, 25.0$^{\prime\prime}$ at 350\,$\mu$m, and 35.7$^{\prime\prime}$ at 500\,$\mu$m.}) at the longer wavelength. Around 15\% of the entries in the catalogue present multiple associations at decreasing wavelength; as a general rule, the closest counterpart was kept without attempting to divide the flux at the longer wavelength among the sources. We assigned to each source the celestial coordinates of the counterpart at the shortest wavelength, where the spatial resolution is higher. Finally, to exclude artifacts, we discarded sources with an axes ratio $>$ 2 and a position angle randomly changing with wavelength by more than 20$^{\circ}$.

\begin{table}[h!]
\caption{\label{tab:tab1} Statistics of the 5\,$\sigma$ \emph{PACS}/\emph{SPIRE} catalogue.}  \small
\begin{center}
\begin{tabular}{cccc}
\hline
   Band                   & \#     &    Sensitivity limit      &         90\% Completeness limit$^a$  \\
                          &        &      (Jy)                 &                (Jy)                  \\
\hline
Total entries$^b$         & 1686   &                           &                           -          \\
70 $\mu$m                 &  658   &	  0.04                 &                 0.21                 \\
160 $\mu$m                &  871   &	  0.09                 &                 0.67                 \\
250 $\mu$m                &  966   &	  0.11                 &                 1.05  (17)           \\
350 $\mu$m                &  697   &	  0.35                 &                 1.32  (22)           \\
500 $\mu$m                &  416   &	  0.46                 &                 1.95                 \\
\hline
\end{tabular}
\end{center}
\begin{list}{}{}
\item[$^{\mathrm{a}}$] For comparison, in parentheses the \emph{BLAST} completeness limit are reported, estimated from Netterfield et al. 2009
(their Figure\,3). 
\item[$^{\mathrm{b}}$] Number of sources detected in at least one band.
\end{list}
\end{table}

\subsection{Catalogue statistics}\label{sec:sec2.2}
The complete \emph{Herschel} catalogue of Vela-C will be published in the next future as part of the HOBYS data products. Here we limit to give in Table \ref{tab:tab1} a statistical summary of the detected sources. In the second column we list the number of entries in each of the five bands: this number increases with increasing wavelength up to $\lambda$\,=\,250\,$\mu$m, than decreases at longer wavelengths. Such behaviour can be explained in the light of the sensitivity limits given in the third column: since the number of entries in the first three bands increases despite the loss of sensitivity, such increasing is a real effect reflecting the intrinsic properties of the cloud. Conversely, in the last two \emph{SPIRE} bands, the poorer sensitivity limits, together with the loss of angular resolution, are likely responsible for the reduced number of the detected sources.  

\begin{figure}
\centering

\includegraphics[width=8cm]{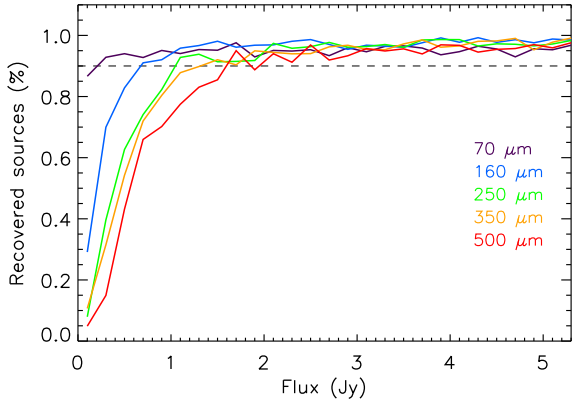}
\caption{Percentage of recovered artificial sources as a function of their fluxes in the five \emph{Herschel} bands.}
\label{compl}
\end{figure}

The flux completeness limits (fourth column) are estimated at the 90\% of confidence level, estimated by recovering artificial Gaussian sources randomly spread over the map and whose fluxes and diameters cover the same ranges found for the observed sources. In Figure\,\ref{compl}, we show the percentage of the recovered artificial sources as a function of the flux for each \emph{Herschel} band.  
For comparison, the \emph{BLAST} completeness limits at 250\,$\mu$m and 350\,$\mu$m (estimated in Figure\,3 of Netterfield et al. 2009) are more than a factor of 15 larger than the \emph{SPIRE} limits at the same wavelengths.

\subsection{Source diameters}\label{sec:sec2.3}
In each of the five bands, the physical FWHM$_{dec}(\lambda)$ has been derived by deconvolving the observed FWHM with the HPBW  at the same wavelength:
\begin{equation}
FWHM_{dec}(\lambda) = \sqrt{FWHM_{\lambda}^2-HPBW_{\lambda}^2}. 
\end{equation}

Noticeably, FWHM$_{dec}(\lambda)$ increases with $\lambda$, with mean values of 7.5$^{\prime\prime}$, 13.4$^{\prime\prime}$, 22.7$^{\prime\prime}$, 28.6$^{\prime\prime}$, 41.0$^{\prime\prime}$ in the five  \emph{Herschel} bands, respectively. Since we are most interested in tracing the cold dust emission (namely that at temperature $\la$ 25 K), which is likely poorly related to the 70\,$\mu$m flux, we define as angular source diameter ($\theta$) the FWHM$_{dec}$ measured at 160\,$\mu$m, where we detect a considerable number of sources with a good angular resolution (see also Motte et al. 2010). For a  small fraction of sources that remained undetected at 160\,$\mu$m, $\theta$ is the FWHM$_{dec}$ measured at 250\,$\mu$m. 

We consider as spatially resolved the detections with $\theta$ larger than $\sim$\,60\% of the angular resolution at 160\,$\mu$m (corresponding to a diameter $D$\,$\sim$ 0.025 pc at the Vela-C distance), as also adopted in the Aquila Rift and Polaris by Andr\'e et al. (2010). 
Noticeably, around $\sim$ 25\% of our sources have a diameter between 0.025 pc and 0.05 pc ('cores'), while most sources are larger ('clumps', $0.5\,\le D\,\le 0.13$ pc); in the following analysis, we will not distinguish any longer between the two categories, and will refer to both of them with the general term 'source'.

\section{Analysis}\label{sec:sec3}

\subsection{Flux scaling}\label{sec:sec3.1}
\begin{figure}
   \centering
 \includegraphics[width=9cm]{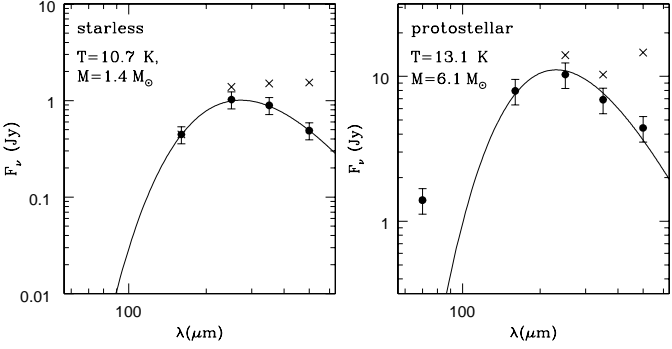}
 \caption{\label{fig:seds} Examples of modified black body fits to the SEDs of two sources (undetected and detected at 70\,$\mu$m, respectively). Crosses and dots mark the data points before and after the correction applied at longer wavelengths to minimize the effects of the poorer spatial resolution (see text). Values of the main fitted parameters are reported.}%
\end{figure}

The physical parameters of the detected sources have been derived by fitting their SEDs with a  modified black body curve. Two examples of the observed SEDs are shown in Figure\,\ref{fig:seds}. These illustrate a trend seen in many of the SEDs, namely that the {\it observed} fluxes (depicted with crosses) flatten or even rise longwards $\lambda \ge$ 250\,$\mu$m.
This effect is a direct consequence of the increase of the FWHM$_{dec}(\lambda)$ with wavelength (Sect.\,\ref{sec:sec2.2}), which in turn implies
increasing areas over which the emission is integrated. In practice, such flattening indicates that a single-temperature, modified black body fit is not completely adequate to model the observed photometric points. However, since the small amount of available data prevents us from using a multiple-temperature, modified black body model, we restricted the fit only to the innermost portions of the sources, whose spatial scales are defined by the source angular diameter $\theta$.
To estimate the emission coming from these restricted solid angles, we followed the flux scaling method adopted by Motte et al. (2010) and described in detail by Nguy$\tilde{\hat{\rm e}}$n Lu{\hskip-0.65mm\small'{}\hskip-0.5mm}o{\hskip-0.65mm\small'{}\hskip-0.5mm}ng et al. (2011). 
This method is based on the idea that for quasi-spherical, self-gravitating sources, the radial density law is $\propto r^{-2}$ (with $r$ $\sim$ 0.1-1 pc), and thus $M(<r) \propto r $ (Shu 1977). This relationship implies that fluxes scale linearly with the source radius, provided that the emission is optically thin and the temperature gradient is weak ($T(r) \propto r^{-q}$, with $q\,\sim 0$, see Motte \& Andr\'e 2001 for more details). 
All these conditions have been empirically verified in our case by taking the photometry at different apertures for some isolated sources, and finding a linear increase of fluxes with increasing radii. Hence, fluxes at 250\,$\mu$m, 350\,$\mu$m, and 500\,$\mu$m, were reduced  
by the ratio FWHM$_{dec}(\lambda)/\theta$. The average values of these factors are 1.9 at 250\,$\mu$m, 2.0 at 350\,$\mu$m, and 3.2 at
500\,$\mu$m, respectively. In Figure\,\ref{fig:seds}, dots represent fluxes after the above correction.

\subsection{Selection criteria}\label{sec:sec3.2}
Once the \emph{SPIRE} fluxes were rescaled, we further filtered the catalogue to select a robust sub-sample for SED analysis. The final
list was composed of sources fulfilling the following criteria, i.e., sources: 
\begin{itemize}
\item[(1)] with detections in at least three adjacent bands between 160\,$\mu$m and 500\,$\mu$m, to contain the peak of the emission of the
cold dust;
\item[(2)] without a dip in the SED between three adjacent wavelengths;  
\item[(3)] not peaking at 500 $\mu$m. Note that conditions (2) and (3) exclude most of the artifacts or detections affected by incorrect positional associations or bad photometry; 
\item[(4)] spatially resolved at 160\,$\mu$m;
\item[(5)] not presenting multiple associations at $\lambda$ $\ge$ 160\,$\mu$m, to alleviate confusion problems; 
\item[(6)] not belonging to the RCW34 region, to exclude sources that could be more distant than 700\,pc (see Sect.\,\ref{sec:sec1}). These sources, identified by eye in the \emph{Herschel} maps, are depicted in yellow in Figure\,\ref{tricro}.
\end{itemize}
The first three criteria are the most stringent ones, since they lead to a selection of only 388 sources out of 1686 present in the catalogue. Applying criteria (4) to (6), we get a final sample of 268 sources to be fitted. Their locations are depicted in Figure \ref{tricro} with different colors indicating their evolutionary stage (see below).

\subsection{SED fitting}\label{sec:sec3.3}
Since cold dust emission is likely unrelated to the photometry measured at 70\,$\mu$m, we fitted the SEDs longward $\lambda$\,$\ge$\,160\,$\mu$m and checked {\it a posteriori} whether or not the modified black body curve fits also the 70\,$\mu$m point.
The assumed relation is:

\begin{equation}\label{eq:flux}
{\rm F}_\nu=(1-e^{-\tau_\mathrm{\nu}})B_\nu(T_d)\Omega
\end{equation}
where F$_\nu$ is the observed flux at the frequency
$\nu$, $\Omega$ is the source area (in sr), B$_\nu$($T_d$) is the black body function at the dust temperature $T_d$, and $\tau_\mathrm{\nu}$ is the optical depth, 
parametrized as $\tau_\mathrm{\nu}=(\nu/\nu_0)^{ \beta}$, where $\nu_0 = c/\lambda_0$ is the frequency at 
which $\tau$=1, and $\beta$ = 2, as predicted by simple dust emission models (Hildebrand 1983). A linear least-squares fit to the scaled fluxes (with their photometric uncertainties) was performed by comparing the observations with a database of models where $T_d$ and $\lambda_0$ are allowed to vary in the following ranges:
8\,K $\le$ $T_d$ $\le$\,40 K; 10\,$\mu$m $\le$ $\lambda_0$ $\le$ 500\,$\mu$m. 
We applied the \emph{PACS} color corrections, amounting to a maximum of 4\% of the flux, to the model fluxes before comparing them with the data (M\"{u}ller et. al 2011).
As a check, we also verified that the fitted values of $\lambda_0$ never exceed 160\,$\mu$m (see Sect.\,\ref{sec:sec4.1} and Figure\,\ref{fig:mass_size}); this insures that the emission is optically thin in the \emph{SPIRE} wavelength range, namely where flux scaling is performed (see the discussion in Sect.\,\ref{sec:sec3.1})

Finally, masses were derived from the dust emission using the relation 
\begin{equation}\label{eq:mass}
M=(d^2 \Omega/k_\mathrm{{300}}) \tau_\mathrm{300}
\end{equation}
where $k_\mathrm{300}$\,=\,0.1 cm$^2$ g$^{-1}$ is the opacity per unit gas mass computed at $\lambda$ = 300\,$\mu$m assuming a gas-to-dust ratio of 100 (Beckwith et al. 1990), and $d$=700 pc. (The details of the derivation of Eq.\,\ref{eq:mass} from Eq.\,\ref{eq:flux} will be given in Pezzuto et al., in preparation).

\subsection{Separating starless from protostellar sources}\label{sec:sec3.4}
Since Vela-C was not observed by \emph{Spitzer} at 24\,$\mu$m, which would have revealed embedded sources with adequate sensitivity, we used the \emph{Herschel} data themselves, and where available, other mid- and far-IR surveys to distinguish starless and protostellar sources. This distinction was primarily done on the basis of the presence of a 70\,$\mu$m counterpart (see also Bontemps et al. 2010). Indeed, a tight correlation between 70\,$\mu$m fluxes and internal luminosities of protostars
has been demonstrated (Dunham et al. 2008), and, furthermore, external heating by the interstellar radiation field cannot 
produce a 70\,$\mu$m emission from a source detectable with \emph{Herschel}.
Hence, we consider as starless all sources without a 70\,$\mu$m counterpart (e.g. Figure\,\ref{fig:seds}, left panel). Conversely, 
protostellar are: {\it i}) sources whose 70\,$\mu$m flux lies above the best modified black body fit by more than 3\,$\sigma$ (e.g. Figure\,\ref{fig:seds}, right panel), or, {\it ii}) sources detected in the mid-IR at $\lambda$ $<$ 70\,$\mu$m by other surveys (IRAS/MSX/Akari). Out of 268 objects, 218 are starless and 48 are protostellar. Two further sources, not detected at $\lambda$ $<$ 70\,$\mu$m by other surveys, were fitted 
with a single modified black body curve from 70\,$\mu$m to 500\,$\mu$m. The temperature of both sources is $\sim$ 25\,K, i.e. warmer than starless sources (Sect.\,\ref{sec:sec3.4}). On this basis we surmise that these sources are embedded Class\,0 protostars (Andr\'e et al. 1993).

\subsection{Physical parameters}\label{sec:sec3.5}
\begin{figure}
   \centering
 \includegraphics[width=9cm]{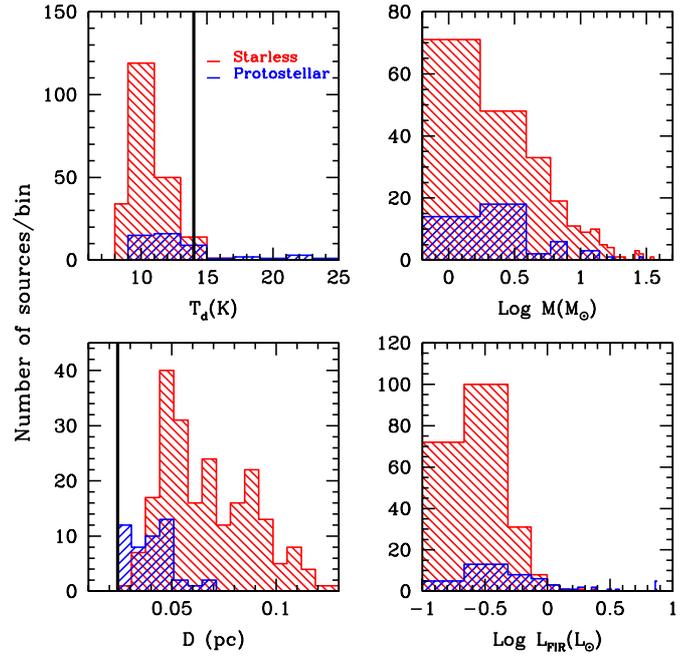}
 \caption{\label{fig:isto} Histograms of the main physical parameters of the sources in Vela-C (top left: temperature; top right: mass; bottom left: diameter; bottom right: \emph{Herschel} luminosity). Starless and protostellar sources are separately highlighted in red and blue, respectively. The black line in the top left panel represents the temperature of the surrounding cloud, taken from Hill et al. (2011), while that in the bottom left panel highlights the spatial resolution limit.}
\end{figure}

In this section, we discuss the statistics of the key physical parameters measured directly from the observations or derived from SEDs fits.
These quantities are listed in Table\,\ref{tab:tab2} and plotted in Figure\,\ref{fig:isto}.

Starless sources exhibit a very narrow temperature distribution (top left panel), strongly peaked at $T_d$ $\sim$ 10 K. The high cutoff is at $T_d \sim$ 15\,K, while the low one coincides with the minimum temperature of 8\,K considered in the fit. On average, protostellar sources are warmer than starless sources; their temperature distribution spreads almost uniformly between 9\,K and 15\,K, though a few sources are found with $T_d$ up to 25\,K. The overall temperature distribution (not shown in Figure\,\ref{fig:isto}) resembles that found from \emph{BLAST} data of the whole Vela region 
(Netterfield et al. 2009). This latter distribution, however, does not show the sharp $\sim$\,10\,K peak, but rather has a shallower shape peaking at $\sim$ 12\,K. This marginal difference may be due to both the better
angular resolution of \emph{Herschel}, which allows deeper penetration into colder regions, and to the inclusion of warmer regions (e.g., Vela-D, Olmi et al. 2009) in the Netterfield et al. sample. 
It is also interesting to compare the compact source distribution with the average temperature of the surrounding cloud. The latter was derived by Hill et al. (2011) by fitting a modified black body to the \emph{Herschel} SED of each pixel. In the parts of the cloud where no compact sources are detected, the temperature is fairly constant at $T_d$\,$\sim$\,14 K (black line in Figure\,\ref{fig:isto}). Practically all the starless sources are colder than the ambient medium, a result expected for dense structures heated only by external radiation.  Interestingly, the same result is found for most of the protostellar sources. This finding reasonably indicates that the external envelope we are tracing is moderately heated by the internal protostar(s), which is therefore likely to be in the very early phases of the evolution.

The mass distribution (top right panel of  Figure\,\ref{fig:isto}) shows a nearly monothonic decrease from lower to higher masses.
There are no significant differences between the average masses of starless and protostellar sources ($\sim$\,5 M$_{\odot}$, Table\,\ref{tab:tab2}); this suggests
that the envelope remains substantially unperturbed by the internal gravitational collapse and, again, that the protostellar sources are relatively unevolved, as little mass accretion is taking place.
Noticeably, 35 sources (of which 7 protostellar and 28 starless) have 10\,$\le$ $M$/M$_\odot$\,$\le$\,20 and 8 (of which 1 protostellar and 7 starless) have 20\,$\le$ $M$/M$_\odot$\,$\le$\,55, whose locations are depicted with big crosses in Figure\,\ref{tricro}. Apart from a few exceptions, these massive objects are rather compact (with diameters between 0.04 pc and 0.07 pc), hence they are potentially able to form individually massive stars.
For comparison, Hill et al. (2011) found in Vela-C nine objects with masses between 13\,M$_{\odot}$ and 70\,M$_{\odot}$. We consider the agreement between the two determinations satisfactory, given the different source extraction methods (Hill et al. 2011 takes into account in the photometry the bias created by bright rimmed effects and source multiplicity) and the different prescriptions under which the SED fitting has been performed (for example, Hill et al. did not apply flux scaling).

The source diameter distributions are presented in the bottom left panel of Figure\,\ref{fig:isto}. For starless sources, it spreads from the resolution limit of 0.025 pc up to 0.13 pc, peaking around 0.05 pc. Such dimensions, as anticipated in Sect.\,\ref{sec:sec2}, span from those of progenitors of single stars, up to those typical of large clumps, that may give rise to the
formation of protostellar clusters. Conversely, protostellar sources are significantly more compact; this suggests that no further fragmentation will occur and that the majority of the protostellar sources likely host single stellar objects or stellar systems with a small number of members.  
The small sizes of protostellar sources can be alternatively (and additionally) explained by considering that the internal source(s) of heating likey causes a temperature gradient and hence a more peaked intensity distribution that is fitted with a compact spatial extent.

Finally, in the bottom right panel of Figure\,\ref{fig:isto}, we plot the distributions of luminosity in the \emph{Herschel} bands ($L_\mathrm{FIR}$). This quantity has been computed by interpolating the corrected fluxes between all available wavelengths between 70\,$\mu$m and 500\,$\mu$m.
For starless sources, $L_\mathrm{FIR}$  is a good approximation of bolometric luminosity; typically, it ranges between 0.05\,L$_\odot$ and 0.5\,L$_\odot$, with a small percentage of objects at larger luminosities not exceeding 5 L$_\odot$.
For protostellar sources, $L_\mathrm{FIR}$ has to be considered as a lower limit to the bolometric luminosity because of the lack of near- and mid-infrared data points in the SEDs.

\begin{table*}
\caption{\label{tab:tab2}. Statistics of the main physical parameters of the sources in Vela-C. We
give median, average, minimum and maximum values.}  \small
\begin{center}
\begin{tabular}{c|ccc|ccc}
\hline
Parameter                          & \multicolumn{3}{c|}{Starless (218)} &  \multicolumn{3}{c}{Protostellar (48)}\\ 
                                   &  median     &  average  & min-max            &  median   &  average    & min-max      \\              
\hline
$M$ (M$_{\odot}$)                  &   3.3       & 5.5       &  0.13-55.8         & 2.7       & 4.8         &  0.15-29.1   \\
$T_d$ (K)                          &   10.0      & 10.3      &  8.0-15.2          & 11.4      & 12.8        &  9.0-24.2    \\
$D$ (pc)                           &   0.064     & 0.067     &  0.025-0.13        & 0.040     & 0.040       &  0.025-0.07  \\
$L_\mathrm{FIR}$ (L$_{\odot}$)     &   0.17      & 0.22      &  0.04-4.8          & 0.6       & 8.0         &  0.08-138    \\
\hline
\end{tabular}
\end{center}
\end{table*}

\section{Discussion}\label{sec:sec4}
\subsection{Virial analysis of starless sources}\label{sec:sec4.1}

\begin{figure}
\centering
\includegraphics[width=8cm]{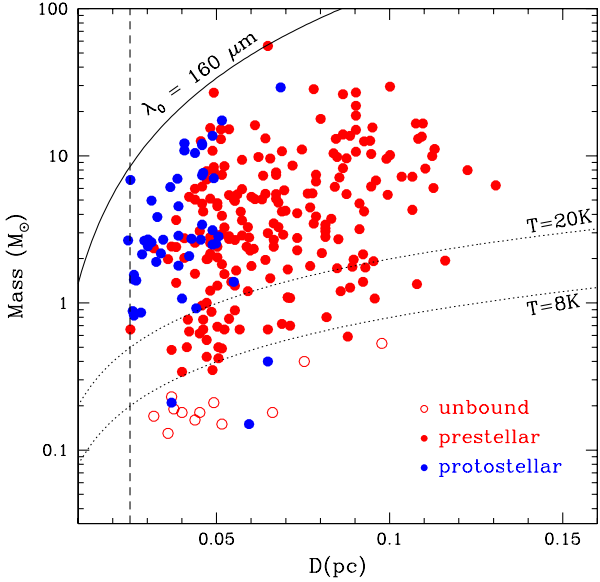}
\caption{\label{fig:mass_size} Mass vs. diameter diagram for the sources in Vela-C. Red filled circles and red open circles are prestellar and unbound sources, respectively. Blue filled circles are protostellar sources. Dotted lines are the {\it loci} of sources with $M$\,=\,0.5 M$_{BE}$, for $T$\,=\,8 K and $T$\,=\,20 K, while the dashed vertical line represents the spatial resolution limit. We also show the curve representing the relation between mass and diameter in the case $\lambda_0$=160\,$\mu$m. This relation shifts towards higher masses with increasing $\lambda_0$: hence, since all our sources are located below the reported curve, they have optically thin emission at $\lambda > 160\,\mu m$.
}
\end{figure}

Starless sources are defined as {\it prestellar} if they are gravitationally bound (e.g., Andr\'e et al. 2000, Di
Francesco et al. 2007), hence they can potentially form one or more stars. In principle, spectroscopic observations would be required in order to derive the virial masses and to securely establish the sources' dynamical states. Since such observations are not available for Vela-C, we have assumed thermal pressure support and neglected the internal turbulence. In this simplified view, the virial mass can be surrogated by the critical Bonnor-Ebert mass:

\begin{equation}
M_\mathrm{BE} \approx 2.4 R_\mathrm{BE} a^2/G
\end{equation} 
Here $a$ is the sound speed at the source temperature\footnote{$a$=$\sqrt{k_B T_d / \mu}$, where k$_B$ is the Boltzmann constant, T$_d$ is the fitted temperature, and $\mu$ = 2.33 m$(H)$ is the mean molecular weight, with m$(H)$ the atomic hydrogen mass.},  $G$ is the gravitational constant, and R$_{BE}$ the Bonnor-Ebert radius (in pc). We take R$_\mathrm{BE}$\,=\,$D$/2.
Starless sources with $M/M_\mathrm{BE}$ $\ge$ 0.5 are selected as being gravitationally bound (Pound \& Blitz 1993) and
candidate prestellar. By this definition, a large number of prestellar sources is found (206 objects), i.e., $\sim$ 94\% of the starless ones, remarkably higher than the percentage (69\%) found in the Aquila Rift cloud (K\"{o}nyves et al. 2010, Andr\'e et al. 2010). Moreover, this percentage mantains high (193 objects, i.e.,\,88\%) even if we demand that bound sources have $M$ at least equal to 1\,$M_\mathrm{BE}$.

To check further this result, we plot all sources in a mass\,vs.\,diameter diagram (Figure\,\ref{fig:mass_size}). First we note that, as a consequence of 
condition (4) in the selection criteria of Sect.\,\ref{sec:sec3.2}, all points lie to the right of the spatial resolution limit of 0.025 pc, shown with a vertical dashed line. Second, we plot the relation between mass and diameter as a function of the second variable in the Equation\,\ref{eq:mass}, namely the wavelength $\lambda_0$ at which the optical depth $\tau$\,=\,1. This relation shifts towards higher masses with increasing $\lambda_0$, therefore all the sources located below (above) the curve for a given value of $\lambda_0$ have optically thin (thick) emission at $\lambda > \lambda_0$.
As example, we show such relation for the value $\lambda_0$\,=160\,$\mu$m, which is the shortest wavelength that we include in the SEDs fits. Since all our sources are located below this curve, their emission is optically thin at wavelengths longer than 160\,$\mu$m. 

Consistently with our virial classification, most of the prestellar sources in Figure\,\ref{fig:mass_size} lie above or between the lines of 0.5 M$_\mathrm{BE}$, computed at 8\,K and 20\,K. We note, however, that a relevant fraction of these objects have a mass up to a factor of ten larger than the Bonnor-Ebert mass at 20\,K. Reasonably, such objects are dynamically unstable, if turbulence and magnetic field supports against gravity are neglected.

All the sources classified as unbound are located below the Bonnor-Ebert curve at $T$\,=\,8 K. Noticeably, most of them cluster around the value $M$ $\sim$ 0.2 M$_{\odot}$. This latter corresponds to the mass detection limit, estimated from the flux sensitivity limits of Table\,\ref{tab:tab1} and by assuming 
average values of $T$ and $\lambda_0$ in Eqs.(2) and (3). Since such mass limit decreases with increasing temperature, we deduce that we are able to efficiently probe only the unbound sources particurlarly warm. Indeed, for the sub-sample of this category (12 objects)  we get $\langle T_d \rangle$=13.8\,K, namely higher than the value of $\langle T_d \rangle$=10.3\,K, referring to the whole sample of starless sources (see Table\,\ref{tab:tab2}). Noticeably, our inability to detect cold, unbound sources, may explain the very large fraction of prestellar sources with respect to the unbound ones (see above). Moreover, this fraction could further decrease if the most massive prestellar cores were gravitationally unstable, as already noticed above.

Finally, to show more clearly the results of Table\,\ref{tab:tab2}, we also plot in the same diagram the positions of protostellar sources. As already noted in  Sect.\ref{sec:sec3.5}, they cluster in Figure\,\ref{fig:mass_size} at diameters smaller than starless sources, and close to our spatial resolution limit. Since no significant differences in mass (as well as in temperature values) are recognizable with respect to starless sources, this result should reflect an increase in average density.

\subsection{Luminosity vs. Mass diagram}

\begin{figure}
   \centering
 \includegraphics[width=9cm]{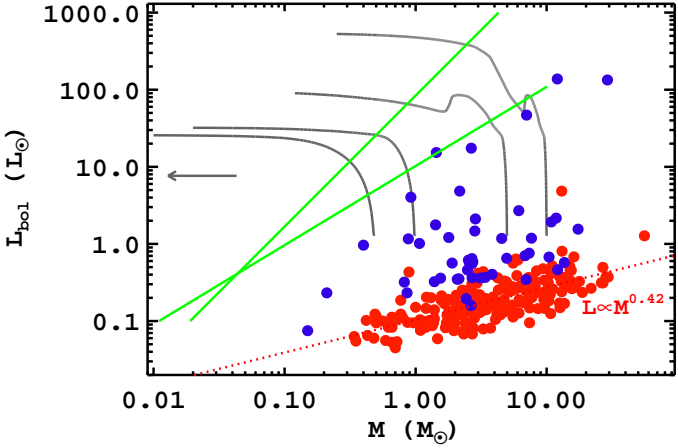}
 \caption{\label{LM} $L_{bol}-M$ diagram for prestellar and protostellar sources in Vela-C (red and blue circles, respectively). Grey solid lines represent the evolutionary tracks for low-mass objects adopted by Molinari et al. (2008), for initial values of 0.5, 1, 5, and 10 M$_{\odot}$, respectively. An arrow indicates the evolution direction, while green lines delimit the region of transition between Class 0 and Class I sources (Andr\'e et al. 2000). The red dotted line represents the best-fitting power law ($L_{bol}\propto M^{0.42\pm0.04}$) for the distribution of the prestellar sources.}
\end{figure}

Another interesting perspective to be examined is the relation between the bolometric luminosities ($L_{bol}$) and the envelope masses of the sources ($M$). To make such comparison, we take as $L_{bol}$ the $L_\mathrm{FIR}$ values, even if these latter well represent bolometric luminosities  only for starless sources. Previous studies (e.g. Saraceno et al. 1996, Andr\'e et al. 2000, Molinari et al. 2008, Elia et al. 2010, Bontemps et al. 2010, Henneman et al. 2010) have illustrated that the $L_{bol}$ vs. $M$ plot is a meaningful tool for  the characterization of the evolutionary status of cores and clumps, since tracks representing model predictions can be directly compared with the locations of the observed sources. In such a diagram, a collapsing core is expected to follow initially an almost vertical path (i.e., mass constant with time, and luminosity strongly increasing with accretion), then subsequently protostellar outflow activity increasingly disperses the envelope while luminosity remains almost constant, resulting in a horizontal track.

In Figure\,\ref{LM}, the $L_{bol}$, $M$ pairs of the Vela-C protostellar and prestellar sources are plotted, along with the evolutionary tracks for low-mass objects adopted by Molinari et al. (2008), based on the simplified assumption of a central protostar accreting mass from
the envelope at a constant rate  $\dot{M}=10^{-5}$ M$_{\odot}$ yr$^{-1}$. More evidently than in previously shown diagrams, a significant distinction between the two populations, although not a complete segregation, is seen. In the same plot, we delimit with green
lines the transition region between Class 0 and Class I objects (Andr\'e et al. 2000), which roughly separates the vertical portions of the tracks, where Class 0 objects are expected to be found, from the horizontal ones, where Class I objects are located. Noticeably, the large majority of the protostellar sources populate the region corresponding to Class 0 objects. This is in contrast with our previous results based on SED fitting, according to which 48 sources are young protostars (although not more evolved) and just two sources are identified as candidate Class 0 sources. Such a discrepancy can be however partly reconciled if considering that our underestimate of the bolometric luminosities of protostellar sources reflects underestimates of the actual ages. Moreover, the Andr\'e et al. (2000) limit separating Class 0 and I could not be completely adequate in Vela, as already found in Taurus by Motte \& Andr\'e (2001). 

Also, the distribution of the prestellar sources is quite flat and homogeneously concentrated in a region of the diagram corresponding to a very early stage where no mass accretion has started.
A tentative power-law fit gives a dependence  $L_{bol}\propto M^{(0.42\pm0.04)}$. This slope is shallower than that found by Brand et al. (2001) for their sample of protostellar sources, namely 0.85. A similar scaling relation has been found for the CO luminosity of clumps: from the theoretical point of view, $L_{CO}\propto M_{vir}^{\sim 1}$ at the virial equilibrium (Scoville \& Sanders 1987; see also Wolfire et al. 1993 and references therein); this slope has been found by Rengarajan (1984), while other authours find lower values (e.g. Solomon et al. 1987,
Yonekura et al. 1997, Miyazaki \& Tsuboi 2000). A direct comparison with our result, however, cannot be performed due to the different classes of sources considered, or different quantities compared.

\subsection{Source mass distribution}
\begin{figure}
   \centering
 \includegraphics[width=9cm]{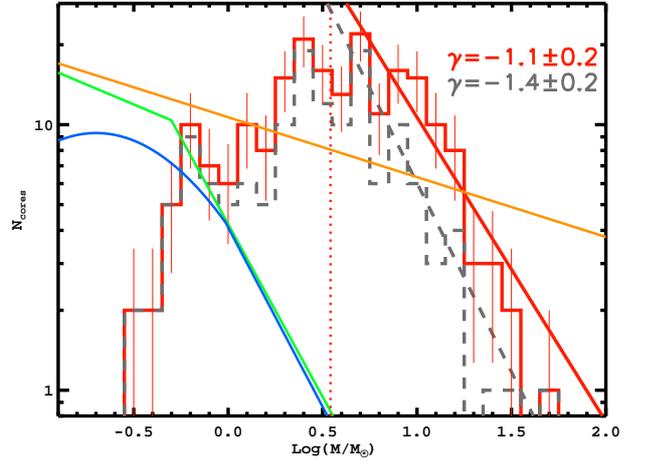}
 \caption{\label{fig:massspectrum} Source mass distribution in Vela-C. The error bars correspond to $\sqrt N$ statistical uncertainties. Continuous red line represents the fit to the linear portion of the mass distribution (namely that consistent with a single straight line within the errorbars) for all the prestellar sources. A similar fit performed on the restricted sample of prestellar sources with diameters less than 0.08\,pc is represented with a grey dashed line. The derived slopes are reported as well.
Mass distribution of CO clumps (Kramer et al. 1998), the single-star IMF (Kroupa 2001) and the multiple-system IMF (Chabrier 2005) are shown for comparison with an orange, a green, and a blue line, respectively. The dotted vertical line represents the mass completeness limit of prestellar sources.}
\end{figure}

Figure\,\ref{fig:massspectrum} shows the mass distribution of prestellar sources in Vela-C.
We estimate a mass completeness
limit of $\sim$ 4 M$_{\odot}$, derived by assuming the average temperature of prestellar sources 
of Table\,\ref{tab:tab2} and the flux 90\% completeness limit at 160\,$\mu$m (see Table\,\ref{tab:tab1}).
Note that this value is significantly better than the \emph{BLAST} completeness limit ($\sim$ 14 M$_{\odot}$ for sources colder than 14\,K). We fit a power law 
$N$(log$M$) $\propto$ $M^{\gamma}$ to the linear portion of the distribution, finding a slope $\gamma$\,=-1.1$\pm$0.2. The uncertainty was determined by considering both the statistical error of the data and the variation of the slope with the histogram binning, which was varied from 0.1 to 0.3 in log($M$/M$_{\odot}$).  
Noticeably, no variations are found in the slope taking into account only the 193 objects for which $M \ge 1 M_\mathrm{BE}$, since their mass distribution differs from that of the whole sample only in the mass range below the completeness limit, where the fit is not performed.
 
The value of $\gamma$ we find in Vela-C is shallower than that found by Netterfield et al. ($\gamma$=-1.9$\pm$0.2, for sources with $T_d<$ 14\,K). If we limit the fit to the \emph{BLAST} mass completeness limit of 14\,$M_{\odot}$, however, we obtain a slope $\gamma\,=-1.9\pm0.2$, which reconciles our high-mass end distribution with that obtained with \emph{BLAST}. More important, and 
thanks to the \emph{Herschel}'s massive increase in sensitivity, which significantly extends the mass range, a possible change of slope is recognizable at $M \ga 10$ M$_{\odot}$. This is close to the value of $\sim 9\,$M$_{\odot}$ where a steepening was probed, for example, in Orion A (Ikeda et al. 2007). Recent theoretical models by Padoan \& Nordlund (2011) predict such change of slope, even if at lower masses ($M\,\sim\,3-5$\,M$_{\odot}$). Unfortunately, no firm conclusion can be drawn on the base of our data because of the poor statistics in the high-mass bins.

Compared with other literature values, our $\gamma$ value of -1.1 is between the mass distribution slope of CO clumps ($\gamma$\,$\sim$-0.7, Kramer et al. 1998) and that typical of prestellar cores, as for example, that measured in the Aquila Rift ($\gamma$\,=\,-1.45$\pm$0.2) on a sample of sources with 2\,$\la\,M$/M$_{\odot}\la\,10$ and diameter typically less than 0.08 pc (K\"{o}nyves et al.\,2010). Hence, our slope reflects the heterogenous nature of our sample, which is composed of objects with a variety of diameters typical of both clumps and cores. Indeed, if we include in the mass distribution fit only sources with diameter less than 0.08 pc (142 objects), we obtain  a slope $\gamma$=-1.4$\pm$0.2 (grey dotted line in Figure\,\ref{fig:massspectrum}), which well reconciles with the slope of the Aquila Rift. Such a slope is also consistent with the IMF slope for 1.0\,$\la\,M$/M$_{\odot}$ (for single-star IMF $\gamma$=-1.3$\pm$0.7, Kroupa 2001, while for multiple-systems IMF $\gamma$=-1.35$\pm$0.3, Chabrier 2005), likely indicating that fragmentation will proceed less efficiently in these small objects.

Finally, a substantial agremeent within the error bars is provided by the comparison with surveys covering a mass range largerly overlapping that of Vela-C. For example, $\gamma\,\sim$\,-1.3$\pm$0.2 for 0.8\,$\,< M$/M$_{\odot}\,<6$ in Perseus, Serpens and Ophiucus (Enoch et al. 2008) and $\gamma$\,=\,-1.3$\pm$0.1 for 3\,$\la\,M$/M$_{\odot}\la\,60$ in Orion A (Ikeda et al. 2007).\\

\section{Conclusions}\label{sec:sec5}
We have reported on the \emph{Herschel} observations of the Vela-C star forming region over an area of $\sim$ 3 squared degrees. From our analysis we obtained the following results:
\begin{itemize}
\item[-] From a 5\,$\sigma$ level catalogue of cold compact sources in five \emph{Herschel} bands between 70\,$\mu$m and 500\,$\mu$m, we have selected a robust sub-sample of 268 sources. Their physical diameters indicate that our sample is mainly composed by 
cloud clumps (0.05 pc $ \la\,D\,\la $ 0.13 pc), together with a $\sim$\,25\% of cores (0.025 pc $ \la\,D\,\la $ 0.05 pc). 
\item[-] Based on the detection of a 70\,$\mu$m flux, we have identified 218 starless and 48 protostellar sources. For two further sources
we do not give a secure classification, but suggest them as candidate Class\,0 protostars.
\item[-] Source physical parameters have been derived from modified black body fits to the SEDs. Both starless and protostellar sources are
on average colder than surrounding medium. This indicates that the radiation from interstellar field and/or from embedded protostars
is unefficient in penetrating the cold dust in deep.
\item[-] Protostellar sources are on average sligthly warmer and more compact than starless sources. Both these evidences can be ascribed to the presence of an internal source(s) of moderate heating, which also causes a temperature gradient and a more peaked intensity distribution. Moreover, the reduced dimensions of protostellar sources may indicate that they will not fragment further.
\item[-] No significant differences are found between the masses of the two groups. These range from sub-solar up to tens of solar masses.
In particular, we find 8 objects with $M >$ 20\,M$_{\odot}$, which are potential candidate progenitors of high-mass stars.
\item[-] More than 90\% of the starless sources result prestellar (i.e. bound) if a virial analysis is applied. This percentage, however, should be considered as an upper limit, both because our sensitivity does not allow us to efficiently probe the coldest unbound sources and because a number of the massive prestellar sources could be gravitationally unstable.
\item[-] A luminosity vs. mass diagram for the two populations of prestellar and protostellar sources has been constructed. Prestellar sources cluster in a well defined region of the diagram corresponding to a very early stage in which no mass accretion
is expected to have started. A tentative power-law fit to the observed distribution gives a dependence $L_{bol}\propto M^{(0.42\pm0.04)}$. Conversely, protostellar sources populate the diagram region corresponding to the early accretion phase.
\item[-] The mass distribution of the prestellar sources with $M \ge$ 4 M$_{\odot}$ shows a slope of $-1.1\pm0.2$. This is between that typical of CO clumps and those of cores in closeby star-forming regions, maybe reflecting the heterogeneous nature of our sample, which is a mixture of cores and clumps. We signal a possible a change of slope in the mass distribution for $M \ga $ 10\,M$_{\odot}$, even if the very big errorbars in the higher mass bins prevent us to draw firm conclusions on its reliability.
\end{itemize}

\begin{acknowledgements}
SPIRE has been developed by a consortium of institutes led by Cardiff Univ. (UK) and including Univ.Lethbridge (Canada); NAOC
(China; CEA, LAM (France); IFSI, Univ.Padua (Italy); IAC (Spain); Stockholm Observatory (Sweden); Imperial College London, RAL, UCL-MSSL, UKATC, Univ. Sussex (UK); Caltech, JPL, NHSC, Univ. Colorado (USA). This development has been supported by national 
funding agencies: CSA (Canada); NAOC (China); CEA, CNES, CNRS (France); ASI (Italy); MCINN (Spain); SNSB (Sweden); STFC and UKSA (UK); and NASA (USA). PACS has been developed by a consortium of institutes led by MPE (Germany) and including UVIE (Austria); KU Leuven, CSL, IMEC (Belgium); CEA, LAM (France); MPIA (Germany); INAF-IFSI/OAA/OAP/OAT, LENS, SISSA (Italy); IAC (Spain). This 
development has been supported by the funding agencies BMVIT (Austria), ESA-PRODEX (Belgium), CEA/CNES (France), DLR (Germany),
ASI/INAF (Italy), and CICYT/MCYT (Spain). Data processing and maps production has been possible thanks to ASI generous support via contracts I/005/07/0-1, I/005/011/0 and I/038/080/0.
\end{acknowledgements}



\end{document}